\documentclass[aps,prl,oneside,twocolumn,showpacs]{revtex4-1}

\usepackage{graphicx}
\usepackage{epsfig}

\usepackage{dcolumn}
\usepackage{bm}
\newcommand{\ignore}[1]{}

\usepackage{color}
\usepackage{amsmath}

\begin{document}
\def\e{\mathcal{E}}

\title{Coherent Perfect Rotation}

\author{Michael Crescimanno, Nathan J. Dawson, James H. Andrews}
\affiliation{Department of Physics and Astronomy, Youngstown State
University, Youngstown, OH 44555-2001}

\date{\today}

\begin{abstract}
Two classes of conservative, linear,
optical rotary effects (optical activity
and Faraday rotation) are distinguished
by their behavior under time reversal. In analogy with
coherent perfect absorption,
where counterpropagating light fields are controllably converted into
other degrees of freedom, we show that only time-odd
(Faraday) rotation is capable of coherent perfect rotation in a linear and conservative
medium, by which we
mean the complete transfer of counterpropagating coherent light fields
into their orthogonal polarization. This
highlights the necessity of time reversal odd processes (not
just absorption) and coherence
in perfect mode conversion and may inform device design.

\end{abstract}


\maketitle

Coherent Perfect Absorption (CPA) \cite{chong10.01, wan11.01}
illuminates the role optical coherence plays in the perfect conversion of
optical energy into other modes (typically incoherent fluorescence or heat).
CPA is a non-conservative linear process,
typically modeled using a non-Hermitian Hamiltonian. In its
original formulation, this non-Hermitian Hamiltonian included absorption/gain
to explicitly break
the time reversal invariance of the underlying fundamental processes.
This is also the case with the formulation of CPA in PT-invariant
theories \cite{longhi10.01,chong11.01}, which has led to a
fertile way to explore
many subtleties in optical processes\cite{longhi11.01,lin11.01}.

\begin{figure}[t]
\includegraphics[width=\linewidth]{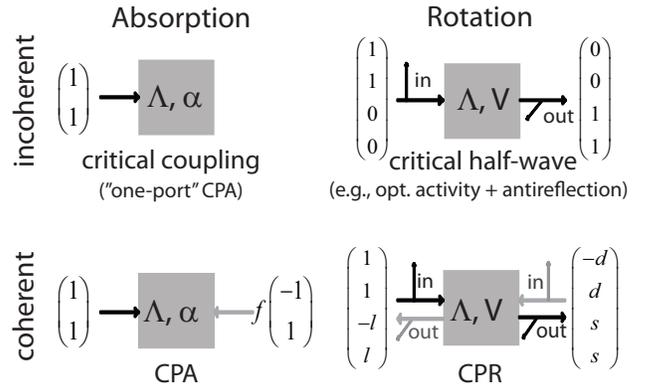}
\caption{CPA and CPR are distinct from critical coupling and
critical rotation\cite{heffe91.01,cai00.01,tisch06.01}.
For a fixed value of $\Lambda$, the system's length
in terms of the vacuum wavelength, critical coupling and
CPA occur at a particular value of the absorption $\alpha$ and index.
Critical half-wave rotation and CPR
first occur at ``threshold'' values for the material's Verdet-magnetic
field product $V$. Only CPA and CPR
depend upon the amplitude and relative phase of the counterpropagating
beams; these can be used to control the onset of CPA or CPR.}
\label{jimfig}
\end{figure}

In this paper we develop theory for Coherent Perfect Rotation (CPR),
the conservative transfer of {\it any} fixed input
polarization state of coherent counterpropagating light fields
completely into its orthogonal polarization.
CPR highlights the necessity of combining
T-odd processes (in, for example, magneto-optics) with optical coherence
to achieve this perfect conversion. By contrast T-even conservative
processes cannot effect such a transformation.
CPR denotes a conservative
(thus fully Hermitian Hamiltonian) process that first appears at
a particular ``threshold'' value of
the parameter scaling the T-odd process, and
there are many
phenomenological correspondences between CPA and CPR,
illustrated schematically in
Fig.  \ref{jimfig}.
Beyond revealing a connection between T-odd processes, Hermiticity, and
CPA, CPR may inform the design of novel magneto-optical sensors
and devices.

We adopt a $4 \times 4$ transfer matrix approach to
describe linear optical transport of a monochromatic
ray moving back and forth along the ${\hat z}$-axis,

\begin{equation}
{\cal M} =
\left( \begin{array}{cc}
M & C \\
B & M' \end{array} \right)
\qquad \mathrm{with} \qquad {\vec v}_{i+1} = {\cal M}_i {\vec v}_i \, ,
\label{M4by4}
\end{equation}
where the $M$'s, $B$ and $C$ are $2 \times 2$ (in general complex) matrices;
here we are
working in the basis where the local field (complex) amplitudes are
${\vec v} = (E_x, H_y, E_y, -H_x)$.
Note that for birefringent materials $M \ne M'$,  but
since we are interested in systems that transform any input polarization
into the orthogonal polarization, we will focus on
the case of non-birefringent materials in which $M = M'$.
The more familiar single polarization form of the transport is
in terms of the $2 \times 2$ $M$ (take $B=C=0$).
Throughout we work in units in which the
familiar propagation eigenstates of a single polarization in the vacuum
are ${\vec e}_R = (E_x, H_y) = (1,1)$ for a rightmoving wave
and ${\vec e}_L = (-1,1)$
for a leftmoving wave.
Thus for review, we represent the coherent scattering from a linear
material whose ($2 \times 2$) transfer
matrix is $M$ by ${\vec e}_{\rm in} = (1,1) + r(-1,1)$
as the incident fields from the left
and ${\vec e}_{\rm out} = t(1,1) = M{\vec e}_{\rm in}$ being the
fields on the right, with $r$ and $t$ denoting the reflection and transmission
amplitudes (generally complex numbers). For
reference, solving the transport in this
basis gives,
$t = 2(m_{11}m_{22} - m_{12} m_{21} ) /(m_{11}+m_{22} -m_{12} - m_{21})$ and
$r = (m_{11} -m_{22} +m_{12} - m_{21})/(m_{11}+m_{22} -m_{12} - m_{21})$, where the
$m_{ij}$ are the matrix elements of $M$
(note difference in basis to Ref.\cite{yeh80.01}).

In the $2 \times 2$ case,
T-symmetry indicates that real diagonal elements of $M$ are T-even
whereas real off-diagonal elements are T-odd.
In general, matrices $C$ and $B$ in ${\cal M}$
can each be written as a sum of
T-even and T-odd parts.
Thus in the chosen basis the T-even part is of the form

\begin{equation}
\left\{C~{\rm or}~B\right\}_{\mathrm{T-even}} =
\left[ \begin{array}{cc}
Re & Im \\
Im & Re\\
\end{array} \right]
\label{Teven_map}
\end{equation}
where $Im$ ($Re$) stand for
imaginary (real) matrix elements.
Note these elements can all be different from one another.

In contrast, for the
T-odd part of the $B$ and $C$ matrices,
\begin{equation}
\{C~{\rm or}~B\}_{\mathrm{T-odd}} =
\left[ \begin{array}{cc}
Im & Re \\
Re & Im\\
\end{array} \right] \, .
\label{Todd_map}
\end{equation}
where, again, all entries could be different.
$T$-odd pieces in the $2 \times 2$ $M$ are associated with
absorption/gain. However, addressing polarization changing
processes in the $4 \times 4$ basis there are
combinations of these T-odd matrix elements that conserve
the total power.

For materials without linear birefringence
the resulting $O(2)$ symmetry about the axial
direction implies
$M=M'$ and
$B=-C$, regardless of the T-symmetry of the underlying matrices.

In steady state, the local power
flux will be a constant of the transport for a conservative system.
A local expression for the power flux in the chosen basis is
$\sim {\vec v}^\dagger {\cal P} {\vec v}$ where ${\cal P} =
 \left[ \begin{array}{cc}
P & 0 \\
0 & P\\
\end{array} \right]$ in which for each polarization
$P =
\left[ \begin{array}{cc}
0 & 1 \\
1 & 0\\
\end{array} \right]$.
The statement that the transport is conservative is thus
${\cal M}^\dagger {\cal P} {\cal M} = {\cal P}$. This leads to important
constraints on the matrix elements of $M$ and $B,C$.
In the more familiar $2 \times 2$ formulation of transport for
a single polarization, a conservative system satisfies $M^\dagger P M = P$,
indicating that  $m_{11}$ and $m_{22}$ must be purely real,
while $m_{12}$ and $m_{21}$ must be purely imaginary (det($M$)=1
is automatic in 1-d linear transport as it preserves the
$E_x, H_y$ commutator)
Note that this is consistent with $M$ being T-even, as expected.
For example, for normal incidence
on a purely dielectric material of thickness $L$, index $n$, the $M =
\left[ \begin{array}{cc}
\cos\delta & {\frac{i}{n}}\sin\delta \\
in\sin\delta & \cos\delta\\
\end{array} \right]$
where $\delta = nk_0L$ and $k_0$ is the vacuum wavenumber.

Similarly, for conservative transport in the full $4 \times 4$ system,
the  $M$, $B$ and $C$ jointly satisfy,
\begin{equation}
M^\dagger P M + C^\dagger P C = M'^\dagger P M' + B^\dagger P B = P
\label{consv1}
\end{equation}
and
\begin{equation}
M^\dagger P C + B^\dagger P M' = 0
\label{consv2}
\end{equation}
The relations
Eqs.(\ref{consv1})-(\ref{consv2}) indicate that
T-symmetry and power conservation are not
identical. Solving Eq.(\ref{consv2})
in the uniaxial case where $M=M'$ and
$B=-C$, if we restrict to T-even, conservative transport,  gives
\begin{equation}
m_{12}c_{21}-m_{21}c_{12} = m_{22}c_{11} - m_{11}c_{22}
\label{}
\end{equation}
along with $c_{21}/c_{12} = m_{21}/m_{12}$ and $c_{11}/c_{22} = m_{11}/m_{22}$.
Combining these equations indicates that $c_{ij} = \alpha m_{ij}$
with $\alpha$ a real constant.
Then, Eq. (\ref{consv1}) gives
$(1+\alpha^2)M^\dagger P M = P$. Then for transport with rotation in a purely
dielectric material, $M
= \cos\gamma
\left[ \begin{array}{cc}
\cos\delta & {{i}\over{n}}\sin\delta \\
in\sin\delta & \cos\delta\\
\end{array} \right]$
and $\alpha = \tan\gamma$.
Thus only a single parameter, $\gamma$ ,  governs
the overall rotation of the frame in the T-even case, as would be the
case for optical activity in which $\gamma$ is proportional to
the product of the concentration of chiral centers
and sample length.

Assuming both that the components of $M$ remain T-even and the system is
uniaxial, the case of conservative, T-odd $C$ in
Eqs.(\ref{consv1})-(\ref{consv2}) reduces to
\begin{equation}
{\rm det}M -{\rm det}C = 1
\label{220}
\end{equation}
and
\begin{equation}
m_{11} c_{22} + m_{22} c_{11} = m_{21} c_{12} + m_{12} c_{21} \, .
\label{220split}
\end{equation}
Thus, studying conjugation and scaling symmetry of the above equations,
we see that there are three (real) parameters that determine the
longitudinal T-odd polarization mixing
in a uniaxial material. One of these parameters is the
ordinary Faraday rotation parameter (the Verdet constant times the applied
longitudinal magnetic field). The
other two parameters in a general solution of Eqs. (\ref{220}) and
(\ref{220split}) are less familiar though lead to the same phenomena.

As emphasized in the literature, the adjective ``coherent'' in CPA and CPR
indicates its reliance on the relative phase between the
counterpropagating light fields in achieving the mode conversion.
Thus CPA and CPR are necessarily two-port processes, in contrast to
critical coupling\cite{heffe91.01,cai00.01,tisch06.01},
itself sometimes referred to as 1-port CPA.
Also in contrast to 1-port devices,
CPR is only possible using T-odd
processes such as Faraday rotation as we now show.

As noted in the original formulation\cite{chong10.01}, CPA can
be understood via $2 \times 2$ transfer matrices, and we
review it briefly here to motivate the $4 \times 4$ transfer matrix
expressions below that describe CPR.
In CPA there are incoming fields only,
and in our choice of basis, these are ${\vec v}_{l} = (1,1)$
and ${\vec v}_{r} = f (-1,1)$ (note $f$ is complex).
These fields are related via the transfer matrix as,
${\vec v}_{r} = M {\vec v}_{l}$ which in terms of the matrix
elements of $M$ indicates that CPA requires the
condition $m_{11} + m_{22} + m_{12} + m_{21} = 0$.
In terms of a fixed optical element size, this (complex)
equation yields both the wavelength of the CPA pole in the S-matrix
and the critical value of the dissipative coupling (which
necessarily has T-odd components in $M$).

It is straightforward to find the location of a CPR resonance using the
$4 \times 4$ basis. For fields on the left take
${\vec v}_l = (1,1,-l, l)$ where $l$ is the amplitude the outgoing
rotated wave. On the right, take ${\vec v}_r =
(-d, d,s,s)$; this configuration thus consists of only incoming
fields of one polarization and  outgoing fields of the orthogonal
polarization only, the CPR state.
In analogy with the CPA state, these boundary conditions
lead to a condition on the
size, wavelength and rotary power of the system.
For uniaxial systems
with the $4 \times 4$  form of ${\cal M}$ as described earlier, we require
\begin{equation}
M \left( \begin{array}{c} 1 \\ 1 \end{array} \right) + C \left( \begin{array}{c} -1 \\
1  \end{array} \right) l
=
 \left( {\begin{array}{c}
-1 \\
1  \end{array}} \right) d
\label{CPRa}
\end{equation}
and
\begin{equation}
-C
\left( \begin{array}{c}
1 \\
1  \end{array} \right)
+ M \left( \begin{array}{c}
-1 \\
1  \end{array} \right) l
=
 \left( \begin{array}{c}
1 \\
1  \end{array} \right) s \, .
\label{CPRb}
\end{equation}
Any optically-active, uniaxial, conservative process never
solves the above pair, and thus cannot be used
to achieve CPR. For this case, as indicated in the preliminaries,
$C \sim M$ and thus
${\cal M} =  \left[ \begin{array}{cc}
M\cos\gamma & -M\sin\gamma  \\
M\sin\gamma  & M\cos\gamma   \end{array} \right]$, for $\gamma$ proportional
to the concentration-length product of the chiral centers.
Using this form in
Eqs.(\ref{CPRa}) and (\ref{CPRb})
and eliminating $l$, $s$ and $d$, we arrive
at the single constraint
\begin{equation}
-(m_{11}-m_{22})^2 + (m_{12}-m_{21})^2 = 4\cos^2\gamma \, ,
\label{constraint}
\end{equation}
Power conservation
discussed earlier indicates
that $m_{11}$ and $m_{22}$ must
be purely real in this basis such that
$m_{12}$ and $m_{21}$ are purely imaginary; thus
Eq. (\ref{constraint}) can never be achieved unless both sides are identically
zero. If so, then
both $m_{11} = m_{22}$ and $m_{12} = m_{21}$. Thus the condition
${\rm det}(M)=1$ would imply that there exists some angle $\phi$ such that
$m_{11} = \cos\phi = m_{22}$
and $m_{12} = i\sin\phi = m_{21}$. For $\phi \ne 0$ then this
case would correspond to a material
that has a net index of refraction of unity.
Alternatively plugging the choice $\phi = 0$ into Eqs.(\ref{CPRa}) and
(\ref{CPRb})
the equations become degenerate,
relaxing the requirement on the index although
yielding a solution for any inputs
($1$ or $d$ in any relation, since $M={\bf 1}$)
independently. This is not CPR; it is instead
the rotation analogue of critical coupling (Fig 1).
To reiterate, such a system
conservatively rotates the polarization of light from any given polarization
state completely into the orthogonal state whether it is illuminated
from one side or the other, independent of any phase relationship
between the incoming fields. Indeed, a single slab of an
optically active material can be tuned in width and chiral concentration
to create this analogue of critical coupling for rotation. There are
likely to be other ways to achieve this rotational analogue of
critical coupling, including one we discuss
below, but again, this is not CPR.

The main new idea of this letter is that
CPR is achievable  with T-odd rotation, as we now show
analytically for a slab dielectric Faraday rotator.
The $M$ and $C$ in the chosen basis for a slab are\cite{kato03.01},
\begin{equation}
M =  {{1}\over{2}} \left[ \begin{array}{cc}
C_1+C_2 & i(S_1/n_1+S_2/n_2)  \\
i(n_1S_1+n_2S_2)  &  C_1 + C_2 \end{array} \right]
\label{Mdielectric}
\end{equation}
and
\begin{equation}
C = {{1}\over{2}} \left[ \begin{array}{cc}
i(C_1-C_2) & -(S_1/n_1-S_2/n_2)  \\
-(n_1S_1-n_2S_2)  &  i(C_1 - C_2) \end{array} \right] \, ,
\label{MnCdielectric}
\end{equation}
where $C_{1,2}$ ($S_{1,2}$) refer to the cosine (sine) of
$\delta_{1,2} = n_{1,2} k_0 L$ in which the $n_1,n_2$ are the indices
of refraction
of the left- and right- circular polarization in the slab, the
$k_0$ refers to the vacuum wavevector and $L$ is the thickness of
the slab. For a dielectric slab in an
external magnetic field pointing along the direction of propagation, the
$\delta n=n_1-n_2$ is proportional to the product of the Verdet and the magnetic
field. Note that
this $C$ given by Eq. (\ref{MnCdielectric}) has
the requisite symmetry of Eq. (\ref{Todd_map})
and is conservative, satisfying Eqs.(\ref{consv1})-(\ref{consv2}).

\begin{figure}[t]
\includegraphics[scale=1]{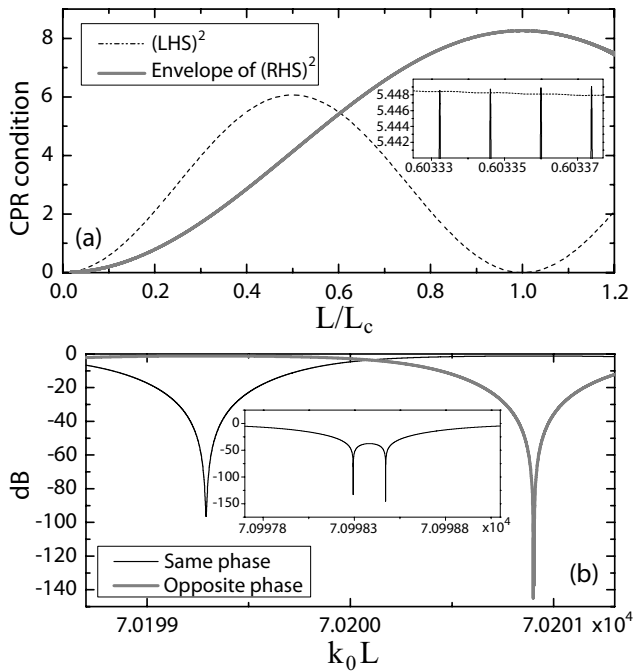}
\caption{(a) The LHS (black dashed line) and the envelope of the fast oscillating RHS (thick gray line) of the CPR condition Eq. (\ref{CPRslab}) plotted as a function of $L/L_c$. The inset is a small portion of the full graph where the fast oscillations of the RHS are shown. (b) Plot of the total reflected intensity in the same polarization as the input fields as a function of the length, $L$, multiplied by the vacuum wave number, $k_0$. The thin line corresponds to the case where the counterpropagating fields have the same relative phase and the thick line to the case where they are 180 degrees out of phase. The inset shows the line splitting of the CPR resonances.}
\label{CPR_point}
\end{figure}

The system
Eqs.(\ref{CPRa}) and (\ref{CPRb}) are 4 (complex) relations for
3 complex quantities ($d, s, l$), so, being overdetermined, demand a
condition on the $n_{1,2}, k_0 L$ which may or may not be physically
satisfiable.
Algebra shows this condition to be
\begin{eqnarray}
& & \left(n_1+\frac{1}{n_1}\right)S_1C_2 - \left(n_2+\frac{1}{n_2}\right)S_2 C_1 = \nonumber \\
& & \pm \left[\left(n_1-\frac{1}{n_1}\right)S_1 - \left(n_2-\frac{1}{n_2}\right)S_2\right] \, .
\label{CPRslab}
\end{eqnarray}
Whenever this condition is satisfied, the fields fall into the
(external) parity eigenstates  $l=\pm s$
and $d=\pm 1$, as expected.
Again, these are necessarily two-port resonances, as is CPA, and thus
examples of CPR states.
A numerical solution is shown in Fig. \ref{CPR_point} for terbium-gallium-garnet with ${\bar n}
= (n_1+n_2)/2 = 1.95$ subject to a coherent $632.8$nm light source and $\delta n = 2.7\times 10^{-5}$ produced by a 1T external field. Here, we have
plotted the LHS-hand-side squared (LHS)$^2$ of
Eq. (\ref{CPRslab}) as a dashed line and the (envelope of the)
right-hand-side squared (RHS)$^2$
as a gray line. The first of many
CPR states exists under these conditions at $L/L_c \approx 0.603$, where $L$ is the length of the slab and $L_c$ is the critical half-wave rotation length.
It is rather easy to understand some general trends in the location of the CPR
resonances in $\lambda$.
Increasing ${\bar n} = (n_1+n_2)/2$ or
$\delta n$ brings the location of the first
CPR resonance into lower $k_0 L$, as would be the case in CPA with
$\delta n$ playing the role of $\alpha$, the absorption constant.
Thus for a fixed $L$ and a given range of $k_0$, there is a threshold
$\delta n$ at which CPR states first appear, again reminiscent of CPA.

Also in Fig. \ref{CPR_point} is a graph of the total power reflected with the
same polarization as the input fields for this case, clearly indicating the first
CPR resonance near $k_0L \approx 70,201$, or $L \approx 7.07$mm. Notably for
this simple slab geometry, all resonances
come in pairs of the same parity, and are part of a
parity-alternating series of pairs of resonances. As in CPA, these
CPR resonances are bound-state like (zero width). Unlike CPA
where there is but one resonance, for CPR
given ${\bar n}$, $L$, and a range of $k_0$ there are many, and
they occur generically in ``doublets.''

Finally, just as one can reach critical coupling in a 1-port version of
CPA, one can see that for particular values of $n_1$, $n_2$, and $k_0L$ there can
be a degeneracy of the positive and negative parity resonances.
For Fig. \ref{CPR_point}, this occurs for $k_0L \approx 116,355$.
At degeneracy, taking linear combinations of the CPR resonances yields
incoherent critical rotation solutions (in detail they are at $S_1 = 0 = S_2$
and $C_1 = -C_2 = \pm 1$). These are optically indistinguishable from
the critical rotator of the optical activity example already discussed.

An experimental verification of CPR is planned using a high Verdet glass.
The CPR resonances are thin, indicating that small changes in a substantial magnetic field
(or in the material itself) may be readily detectable through changes in the
extinction of a reflected polarization. At the level of technological
application, note that an optical modulator based on CPA
will necessarily have limited
dynamic range as the material will always absorb some of the light
even when not in CPA. A CPR-based optical modulator may not suffer
the same limitations.

In conclusion, we have shown that Faraday rotation has the appropriate
symmetries to manifest Coherent Perfect
Rotation (CPR) and analytically developed an example of CPR in
a dielectric Faraday slab rotator. CPR has deep phenomenologically similarity
with CPA, but with a Hermitian Hamiltonian. It appears likely that other
types of coherent perfect mode conversion will have similar phenomenology,
and necessitate T-odd processes.

\section*{Acknowledgments}
The authors are grateful to the National Science Foundation for
financial support from the Science and Technology Center
for Layered Polymeric Systems under grant number No. DMR 0423914 and
to the Ohio
Third Frontier Commission for support of
the Research Cluster on Surfaces in Advanced Materials.

\end{document}